\documentclass[aps,prb,10pt,twocolumn]{revtex4-1}
\usepackage{amsmath}
\usepackage{amsfonts}
\usepackage{amssymb}
\usepackage{bm}
\usepackage{graphicx}
\usepackage[colorlinks=true,linktocpage=true,breaklinks=true]{hyperref}

\begin{document}

\title{Modeling Dzyaloshinskii-Moriya Interaction at Transition Metal Interfaces: Constrained Moment versus Generalized Bloch Theorem.}
\author{Yao-Jun Dong}
\email{yj$\_$dong@outlook.com}
\author{Abderrezak Belabbes}
\email{abderrezak.belabbes@kaust.edu.sa}
\author{Aur\'elien Manchon}
\email{aurelien.manchon@kaust.edu.sa}
\affiliation{King Abdullah University of Science and Technology (KAUST), Physical Science and Engineering Division (PSE), Thuwal 23955-6900, Saudi Arabia}


\begin{abstract}
Dzyaloshinskii-Moriya interaction (DMI) at Pt/Co interfaces is investigated theoretically using two different first principles methods. The first one uses the constrained moment method to build a spin spiral in real space, while the second method uses the generalized Bloch theorem approach to construct a spin spiral in reciprocal space. We show that although the two methods produce an overall similar total DMI energy, the dependence of DMI as a function of the spin spiral wavelength is dramatically different. We suggest that long-range magnetic interactions, that determine itinerant magnetism in transition metals, are responsible for this discrepancy. We conclude that the generalized Bloch theorem approach is more adapted to model DMI in transition metal systems, where magnetism is delocalized, while the constrained moment approach is mostly applicable to weak or insulating magnets, where magnetism is localized.
\end{abstract}

\maketitle


\section{Introduction}

Magnets lacking inversion symmetry in their bulk or at interfaces are currently attracting a growing amount of interest due to the emergence of an antisymmetric exchange interaction, called Dzyaloshinskii-Moriya interaction \cite{Dzyaloshinskii1957, Moriya1960} (DMI). DMI lies at the heart of a vast endeavor aiming at the achievement of novel exotic magnetic phases, ranging from homochiral magnetic spirals \cite{Heide2009} and N\'eel domain walls \cite{Thiaville2012} to magnetic skyrmions \cite{Bogdanov2001,Roszler2006}. These various magnetic textures display intriguing dynamical properties that make them particularly suitable for spintronics operation: Chiral spin spirals can convey information \cite{Menzel2012}, N\'eel walls have been reported to move as fast as several hundreds of meter per second under the action of spin-orbit torques \cite{Miron2011,Yang2015a}, while magnetic skyrmions\cite{Chen2015b,Jiang2015,Woo2016,Moreau-Luchaire2016, Boulle2016b} have been proposed to move robustly against certain classes of pinning potentials\cite{Fert2013,Iwasaki2013b,Sampaio2013}. 

This recent interest for non-centrosymmetric magnetic multilayers has renewed the studies on DMI and several experimental techniques aiming at accurately determining this interaction have been developed lately. At transition metal interfaces, DMI has been determined using spin-polarized scanning tunneling microscopy\cite{Bode2007,Ferriani2008,Meckler2009,Menzel2012}, Brillouin light scattering \cite{Di2015,Nembach2015}, scanning nitrogen vacancy center magnetometry \cite{Tetienne2015,Gross2016}, propagative spin wave spectroscopy \cite{Lee2016c}, spin-polarized low-energy electron microscopy, \cite{Chen2013c,Chen2013d}, magneto-optical imaging of field-induced chiral magnetic bubble expansion \cite{Je2013,Hrabec2014,Pizzini2014}, asymmetric magnetic hysteresis\cite{Han2016} or current-driven domain wall motion \cite{Emori2013,Ryu2013}. These various techniques have provided numerous estimations of DMI strength at metallic interfaces, yielding values up to 2 mJ/m$^2$ depending on the heavy metal/ferromagnet combination.

From the theoretical perspective, DMI has been originally investigated in Mott insulators\cite{Dzyaloshinskii1957, Moriya1960}, such as $\alpha$-Fe$_2$O$_3$ and CrF, where antiferromagnetism arises from superexchange between localized magnetic orbitals\cite{Anderson1959}. It was subsequently extended to weak ferromagnets \cite{Fert1980}, such as Pt- or Au-doped CuMn alloys, where magnetic ions are embedded in a sea of conduction electrons that convey the magnetic exchange through Ruderman-Kittel-Kasuya-Yoshida (RKKY) coupling. In the presence of metallic impurities with spin-orbit coupling (SOC), these conduction electrons experience spin-dependent scattering that gives rise to an antisymmetric magnetic exchange. Interestingly, Moriya's and Fert's theories, although involving quite different orbitals and transport mechanisms, both produce the same form of interaction energy, $E_{\rm DM}=\sum_{ij}{\bf D}_{ij}\cdot({\bf S}_i\times{\bf S}_j)$, as predicted by Dzyaloshinskii based on symmetry arguments. Here ${\bf S}_{i}$ is the magnetic moment at position $i$, and ${\bf D}_{ij}$ is the so-called DM vector. Although DMI has been proposed at metallic interfaces quite early \cite{Fert1990}, its physical origin in transition metal multilayers has only been addressed very recently\cite{Heide2008,Heide2009}, triggered by the observation of chiral spin spirals at Mn/W(110) surfaces\cite{Bode2007}. Modeling DMI at transition metal interfaces requires the proper treatment of the delocalized character of the $d$ orbitals, a feature only accessible through numerical (tight-binding or density function theory - DFT) techniques. Several theoretical approaches have been proposed to evaluate DMI using first principles calculations. \par

Because of the cross product ${\bf S}_i\times{\bf S}_j$, DMI is antisymmetric upon $i\leftrightarrow j$ permutation. Therefore, two spin spirals with opposite sense of rotation, i.e. different chiralities, display a shift in energy that can be related to the DM vector. This effect resembles the influence of Rashba coupling on itinerant electrons: it distorts the energy dispersion \cite{Costa2010}. The main technical difficulty with computing DMI using DFT is to accurately evaluate the energy of the system in the presence of both spin-spiral (i.e. non-collinear magnetism) and SOC (the spin is not a good quantum number). A successful approach to compute the energy shift is to build spin spirals in the reciprocal space \cite{Kurz2004} employing the generalized Bloch theorem \cite{Herring1966,Sandratskii1991} in the absence of SOC. DMI is then computed to the {\em first order} in SOC \cite{Heide2008,Heide2009}, which limits this method to materials with weak enough SOC. This approach has been successfully used to compute DMI in a wide range of transition metal interfaces\cite{Kashid2014,Zimmermann2014,Dupe2014,Dupe2016,Belabbes2016,Belabbes2016b}. Following the same spirit, Freimuth et al. \cite{Freimuth2013b,Freimuth2014,Freimuth2017} and Kikuchi et al. \cite{Kikuchi2016} developed a linear response theory that provides an estimation of the DMI in the limit $q\rightarrow0$ and related it to the Berry curvature in the mixed spin-momentum space. In Kikuchi's theory, DMI is expressed as  $E_{\rm DM}=(\hbar/2)\int_\Omega {\bf m}\cdot[({\cal J}_s\cdot{\bm\nabla})\times{\bf m}]d^3{\bf r}$, where ${\cal J}_s$ is the equilibrium spin current that interacts with the magnetic texture. Mankovsky and Ebert\cite{Mankovsky2017} have recently computed DMI using Freimuth's theory implemented on fully relativistic Korringa-Kohn-Rostoker Green's function technique. \par

Alternatively, a method that does not require the construction of spin spiral in reciprocal space has been recently proposed. It consists in building a real space spin cycloid in the presence of SOC by constraining the direction of the magnetic moments\cite{Constrained}, and extract the DM vector based on a micromagnetic model; the energy difference between two counterrotating spin spirals is proportional to the DM vector. This approach has been used to compute DMI in ferroelectric magnets such as MgCr$_2$O$_4$\cite{Xiang2011} or Cu$_2$OSeO$_3$\cite{Yang2012b} and recently extended to transition metal interfaces\cite{Yang2015c}. The "constrained moment" method has the advantage of being applicable to materials with large SOC, but becomes computationally prohibitive in the long-wavelength limit (i.e. when the spin spiral exceeds 10 atoms). \par

An important question that has not been addressed till now is how comparable these different methods are. We are particularly interested in confronting the results obtained by building spin spirals in reciprocal space ("generalized Bloch theorem" approach) to the ones obtained using spin spirals in real space ("constrained moment" approach). In the present work, we compute the DMI energy in Co/Pt bilayers using these two methods for a large range of nearest-neighbor rotation angles (i.e. wavelengths). We find that although the magnitude of the DM energy is comparable, its dependence as a function of the spin spiral wavelength is very different in the two methods. This study points out the limit of applicability of the "constrained moment" approach, emphasizing the importance of long-range magnetic interactions in transition metal interfaces.


\section{Computational Method\label{s:method}}

\begin{figure}[h]
    \centering
    \includegraphics[scale=0.3]{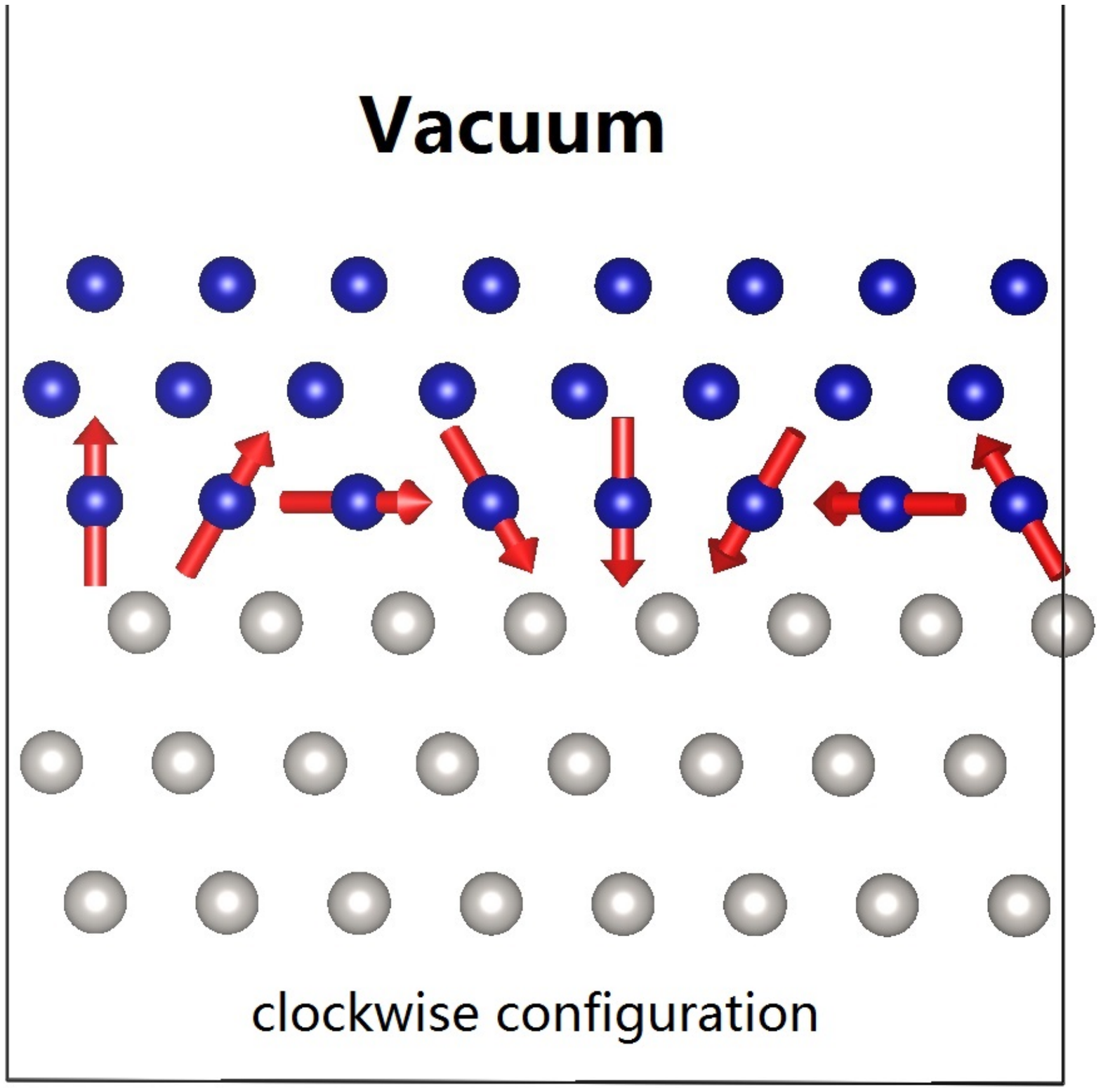}
    \caption{(Color online) Profile view of $8\times1$ hcp(0001)Co/fcc(111)Pt bilayer structure with clockwise spin configuration. The corresponding angle between neighboring moments is 45$^\circ$.}
    \label{fig0}
\end{figure}

\subsection{Method 1: Constrained Moment}

The first method we employ consists in building a spin cycloid in real space on a $n\times1$ supercell, where $n$ is the length of the cycloid, as displayed on Fig. \ref{fig0}, and estimating the energy shift obtained upon switching the chirality of the cycloid\cite{Yang2015c}. For the sake of comparativeness, we closely follow the procedure proposed by Yang et al.\cite{Yang2015c}. The supercell is composed of 1 to 3 monolayers of Co deposited on 1 to 7 monolayers of noble metal (either Pt or Au), and computed using the VASP package, within the Perdew-Burke-Ernzerhof generalized gradient approximation \cite{Perdew1996} of the exchange correlation interaction \cite{Kresse1993, Kresse1996}. An energy cutoff of 320 eV is selected for the plane wave expansion. First, the geometry is relaxed until the force acting on each atom becomes less than 0.005 eV/\AA. The self-consistent total energy of hcp(0001)Co/fcc(111)Pt structures is calculated by imposing the magnetic moment configuration using the constrained moment method implemented in VASP \cite{Constrained}. A penalty functional is added to the total energy in order to align the magnetic moment along a preferred direction. In the present work, we tried to keep this penalty contribution as small as possible, in order to reduce its influence on the total energy of the system. The minimal penalty energy ranges from 2.9 to 3.5 meV for Pt(X)/Co(3) bilayers. In this notation, the number in parenthesis refers to the number of monolayers.\par

The total DMI energy $d^{tot}$ can be estimated from the energy difference of two spin spirals with different chiralities,\cite{Yang2015c}
\begin{eqnarray}
\Delta E_{\rm DM} = E_{\rm CW} - E_{\rm ACW} = 3 n d^{tot} \sin\frac{2\pi}{n} ,
\label{eq1}
\end{eqnarray}
where $E_{\rm CW}$ and $E_{\rm ACW}$ are the total energy of clockwise and anti-clockwise configurations. This expression assumes nearest-neighbor exchange interaction only, i.e. ${\bf D}_{ij}\neq0$ only if sites $i$ and $j$ are nearest-neighbors. The surface energy density $D$ reads\cite{Yang2015c}
\begin{eqnarray}
D = \frac{3\sqrt{2}{d^{tot}}}{N_Fa^2},
\label{eq2}
\end{eqnarray}
where $a$ is the fcc lattice constant and $N_F$ is the number of magnetic overlayers. \par

\subsection{Method 2: Generalized Bloch Theorem}

The second method consists in computing the DMI energy using the generalized Bloch theorem\cite{Kurz2004}, as described in Refs. \onlinecite{Belabbes2016,Belabbes2016b}. This is done by performing DFT calculations in the local density approximation\cite{Perdew1981} to the exchange correlation functional, using the full potential linearized augmented plane wave method in film geometry\cite{Wimmer1981} as implemented in the FLEUR code\cite{Fleur}. We considered 512 k-points in the two-dimensional Brillouin zone for scalar-relativistic calculations and 1024 k-points for the calculation with SOC treated within first-order perturbation theory. In order to investigate the DMI, we first self-consistently calculate the total-energy $E({\bf q})$ of planar (N\'eel) spin spirals employing the generalized Bloch theorem within the scalar-relativistic approach\cite{Heide2008}. The spin spirals are defined at position ${\bf r}_i$ by the magnetic moment direction ${\bf m}_i = (\cos{\bf q}\cdot{\bf r}_i, \sin{\bf q}\cdot{\bf r}_i,  0)$, where {\bf q} is the propagation wave vector. Varying the ${\bf q}$-vector in small steps along the paths connecting the high-symmetry points of the two dimensional Brillouin zone, we find the well-defined magnetic phases of the hexagonal lattice: ferromagnetic state at $\bar{\Gamma}$- point ($q$ = 0), antiferromagnetic state at $\bar{\rm M}$-point, and periodic 120$^{\rm o}$ N\'eel state at $\bar{\rm K}$-point. Finally, we evaluate the DMI contribution from the energy dispersion of spin spirals by applying the SOC treated within first-order perturbation theory\cite{Wimmer1981,Heide2009}. 


\section{Results}

\subsection{Benchmark}

\begin{figure}[h]
    \centering
    \includegraphics[scale=0.35]{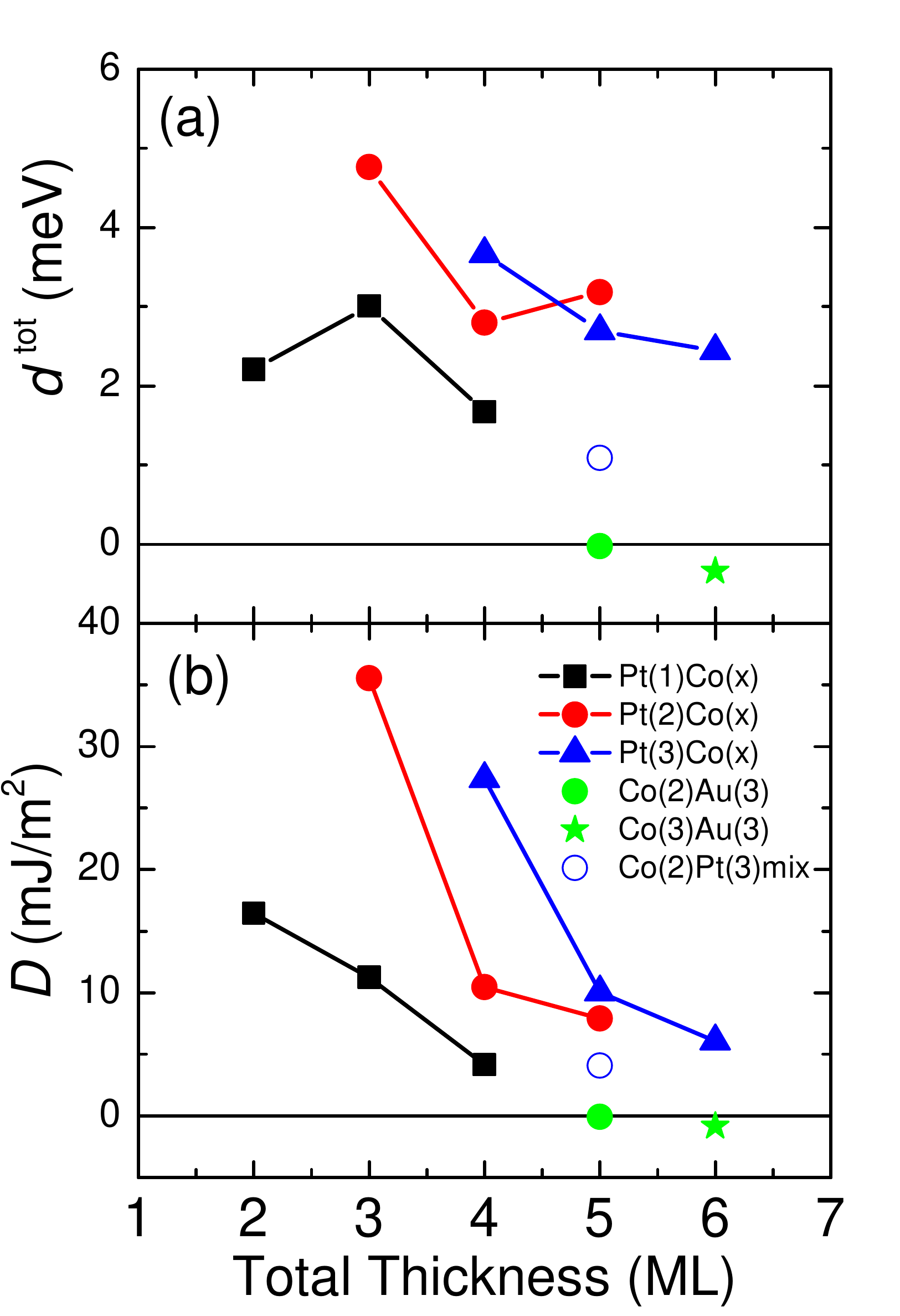}
    \caption{(Color online) (a) Total DMI energy $d^{tot}$ and (b) DMI surface energy density for various interface configurations, as a function of the total number of atomic layers.}
    \label{fig1}
\end{figure}

Let us first benchmark our "constrained moment" calculations to the ones reported by Yang et al.\cite{Yang2015c}. The results for a cycloid of wavelength $n=4$ (which corresponds to an angle of 90$^{\rm o}$ between neighboring moments) are reported on Fig. \ref{fig1}. The total DMI energy and surface energy density are reported for various interfaces such as Co/Pt, Co/Au, and mixed interfacial Co/Pt bilayers. The values we obtained are in agreement with Ref. \onlinecite{Yang2015c}, Figure 2, up to a multiplicative factor of $\sqrt{3}$ as explained in Ref. \onlinecite{Yang2015g}, yielding a total DMI strength $d^{tot}$ for Co/Pt bilayers ranging from 1.5 to 5 meV. The DMI surface energy density $D$ decreases upon increasing the Co thickness, as expected from the interfacial origin of this interaction. The value of DMI is further reduced by half when considering intermixing, i.e. one Pt atom is replaced by one Co atom at the interface (see blue hollow circles in Fig. \ref{fig1}). These results are overall in good agreement with Yang et al.\cite{Yang2015c}, indicating that our procedure is comparable to theirs.

\subsection{Role of substrate thickness}

\begin{figure}[h]
    \centering
    \includegraphics[scale=0.4]{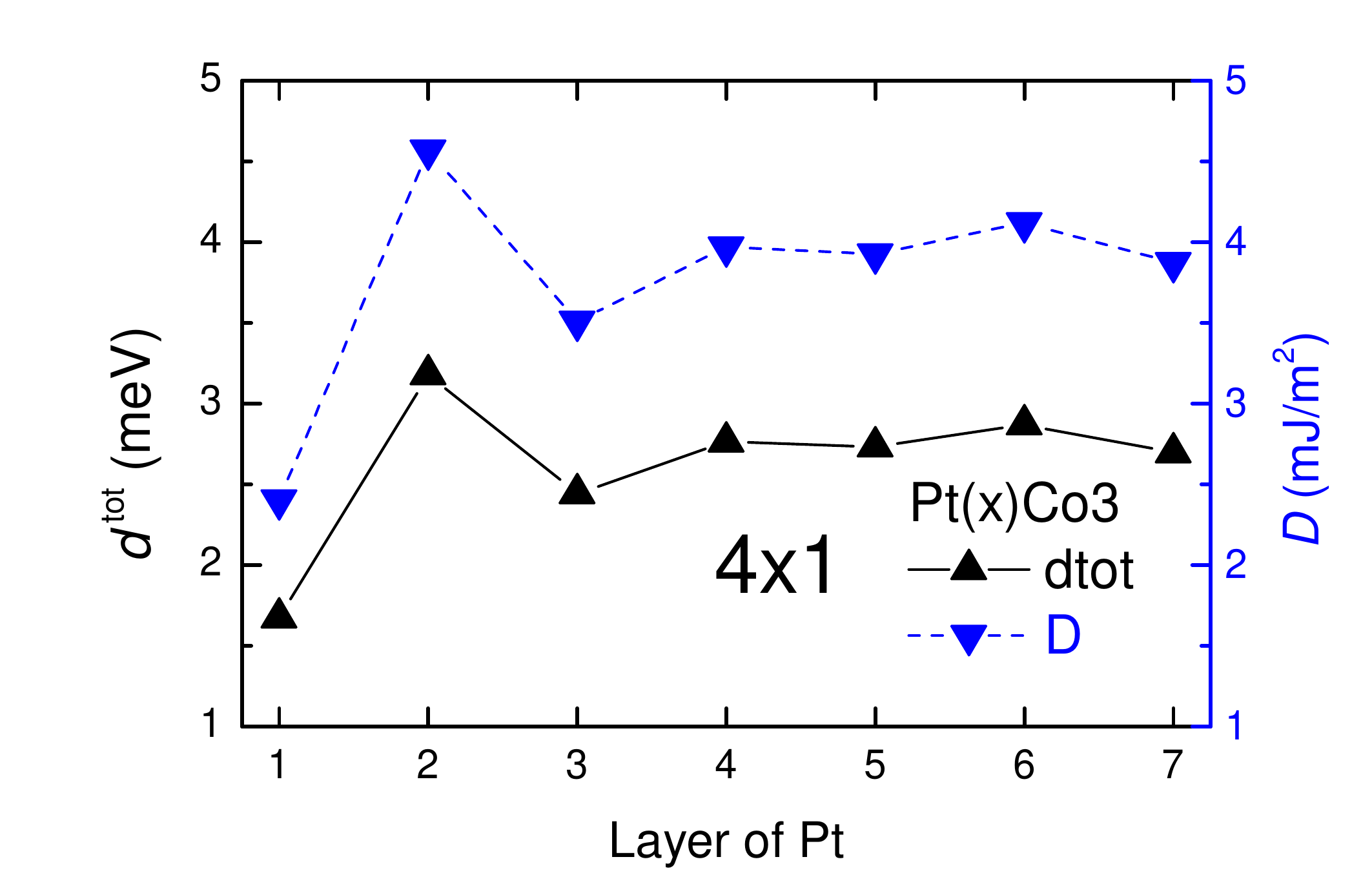}
    \caption{(Color online) Total DMI energy $ d^{tot} $ and surface energy density $D$ for Pt(X)/Co(3), calculated for a cycloid of wavelength $n$=4.}
    \label{fig2}
\end{figure}

While searching for the phenomenological origin of DMI, Ryu et al. \cite{Ryu2013,Ryu2014} suggested that the induced magnetization in the 5$d$ substrate might play an important role on the onset of this interface. Indeed, 5$d$ elements - and in particular Pt - have the ability to become spin-polarized by proximity with an adjacent ferromagnet (see Refs. \onlinecite{Grytsyuk2016}, \onlinecite{Blugel2007} and references therein). Hence, is it reasonable to expect that interfacial properties, such as DMI and magnetic anisotropy, get affected by the induced magnetization. Actually, no evidence of such a correlation has been found in computational studies\cite{Yang2015c,Belabbes2016}. Nevertheless, Tacchi et al.\cite{Tacchi2017} recently reported a progressive buildup of DMI energy upon varying the substrate thickness in Pt/CoFeB, followed by a saturation after 2 nm (i.e. 5 monolayers of Pt). To understand this behavior, the dependence of both $d^{tot}$ and $D$ as a function of the number of Pt monolayers is reported on Fig. \ref{fig2}. We find that after a progressive buildup, both $d^{tot}$ and $D$ reach a constant value around 4 monolayers of Pt.

\begin{figure}[h]
    \centering
    \includegraphics[scale=.3]{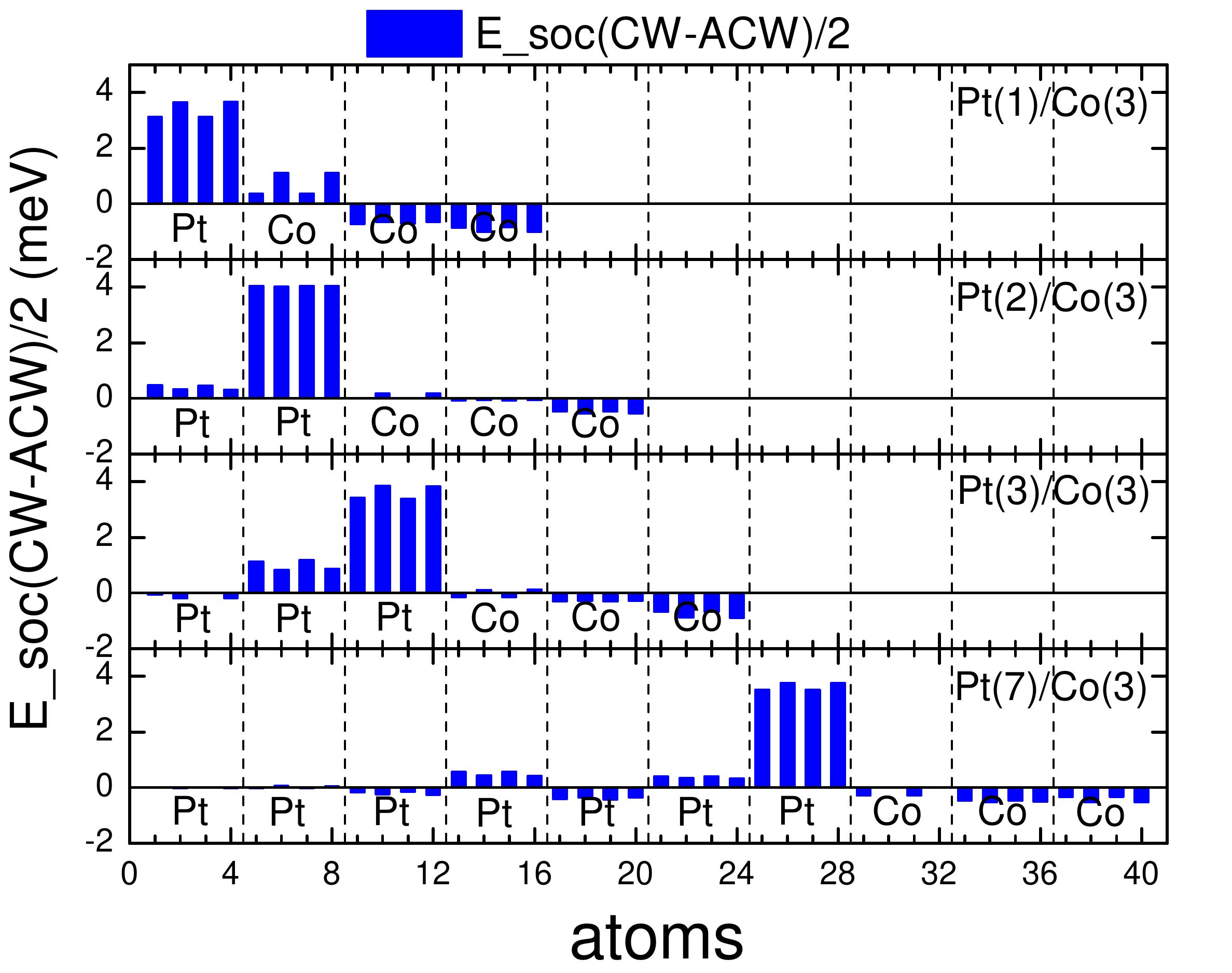}
    \caption{(Color online) $\Delta E_{\rm DM}$ of each atomic site for different Pt(X)/Co(3) interfaces. The monolayers are indicated by the vertical dashed line, each monolayer being composed of 4 atoms.}
    \label{fig3}
\end{figure}

To further investigate the origin of this effect, we compute the layer-resolved contributions to $\Delta E_{\rm DM}$ (the energy difference between clockwise and anti-clockwise cycloids) for a $4\times1$ supercell composed of 3 Co monolayers deposited on $m$ Pt monolayers ($m$=1, 2, 3, and 7). The results are reported on Fig. \ref{fig3}. In this figure, each vertical bar corresponds to the contribution of one atom, each monolayer comprising 4 atoms. It is clear that the topmost Pt monolayer provides the largest contribution to DMI due to the coexistence of large SOC, interfacial symmetry breaking and induced magnetization. When increasing the substrate thickness, we note that the additional Pt monolayers also contribute to DMI, to a lesser extend and up to 4 monolayers away from the interface. This behavior reflects the spatial dependence of the induced magnetization (which can extend up to 4 monolayers away from the interface in Pt/Co\cite{Grytsyuk2016}) and of the interface-induced symmetry broken character of the wave function. Such a spatial extension of the contribution to DMI in the substrate has also been observed in W(7)/Mn(1), up to 6 monolayers away from the interface \cite{Belabbes2016}. This study emphasizes the {\em non-local character} of DMI at transition metal interfaces (see discussion below in Section \ref{s:discu}) and indicates that thick enough substrate should be taken into account when performing calculations on transition metal interfaces. For the sake of comparison, Yang et al. \cite{Yang2015c} performed calculations on 3 monolayers of Pt, i.e. at the limit of saturation for $d^{tot}$ and $D$, while Belabbes et al.\cite{Belabbes2016} used 6 monolayers and Freimuth et al.\cite{Freimuth2014,Freimuth2014a} used 10 monolayers. Thus, the thicker the substrate, the more accurate the result is. Notice that the three Co overlayers also contribute to DMI but to a much smaller extent (less than 10\% overall). Their contribution is larger in the case of Pt(1)/Co(3) than in Pt(7)/Co(3), probably due to quantum confinement.

\subsection{Constrained moment versus generalized Bloch theorem}

An important question that remains to be addressed is the dependence of DMI as a function of the cycloid wavelength. Indeed, the characteristic length of a magnetic texture depends on the competition between magnetic exchange, anisotropy and DMI energies. The wavelength of the ground state spin spirals observed at transition metal interfaces ranges from 4 to 8 nm \cite{Bode2007,Ferriani2008,Menzel2012}, while the radius of metastable skyrmions in similar systems ranges from 50\cite{Woo2016,Moreau-Luchaire2016} to 200 nm\cite{Boulle2016b}. A robust method to compute DMI should therefore provide consistent values for all these various textures. 

\begin{figure}[h]
    \centering
    \includegraphics[scale=0.3]{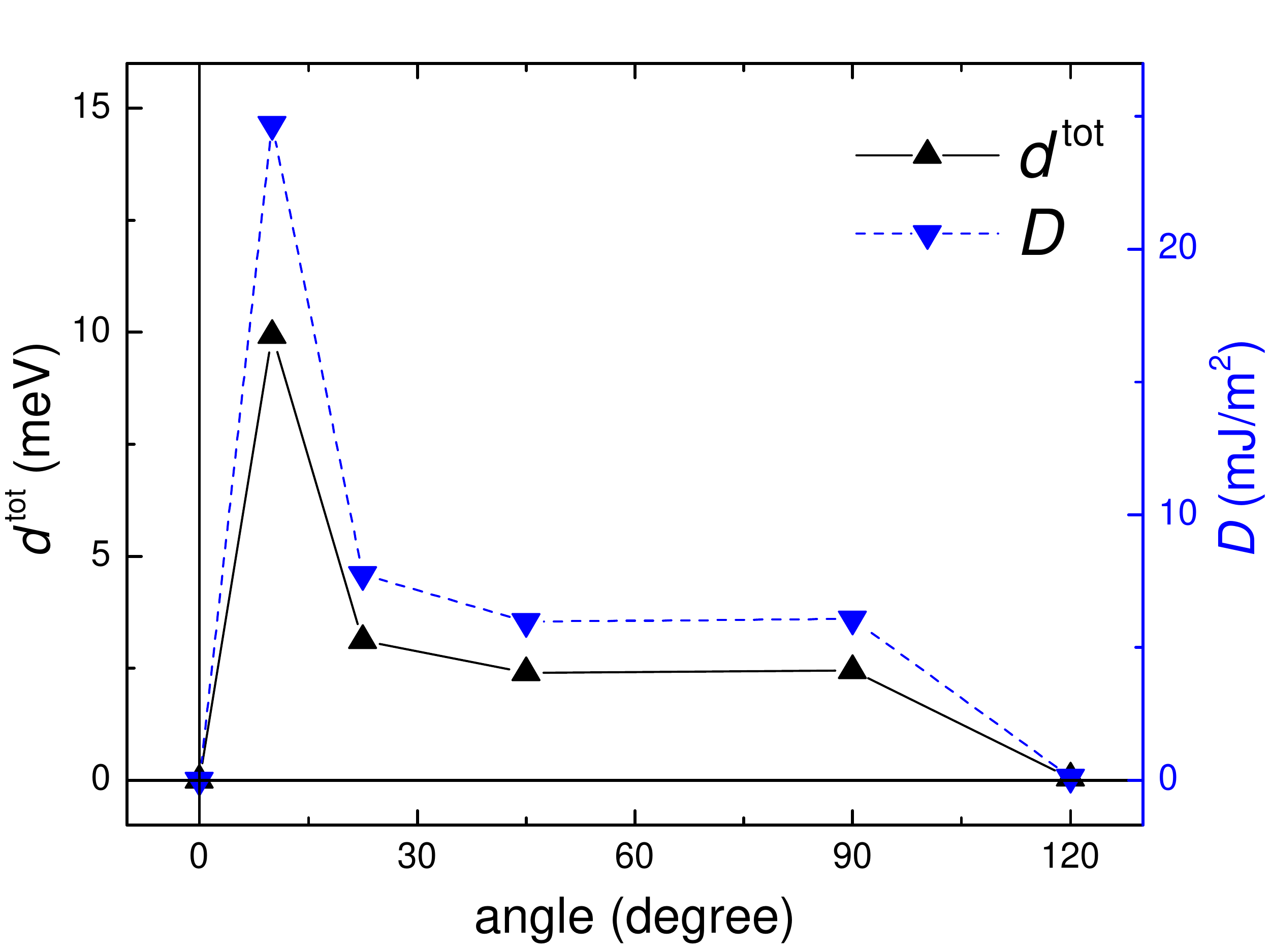}
    \caption{(Color online) DMI energies, $d^{tot} $ and $D$, for Pt(3)/Co(3) with different wavelengths. While the DMI value is unchanged for rotation angles between 22.5$^\circ$ and 90$^\circ$, a large enhancement is found at 10$^\circ$ and a quenching is obtained at 120$^\circ$.}.
    \label{fig4}
\end{figure}

In their study, Yang et al.\cite{Yang2015c} verified that DMI remains unchanged (within 7\%) when increasing the spiral period from $n$=6 to $n$=8. In the present work, we expanded the computation to $ n\times1$ supercells, with a spiral period of $n$ =3, 8, 16, and 36, corresponding to an angle of rotation between neighboring moments of 120$^{\circ} $, 45$^{\circ} $, 22.5$^{\circ} $, and 10$^{\circ}$, respectively. The results are reported on Fig. \ref{fig4} for both $d^{tot}$ and $D$. While little variation is observed between $n$=4 (90$^\circ$) and $n$=16 (22.5$^{\circ}$), in agreement with Ref. \onlinecite{Yang2015c}, the DMI energy changes dramatically for much larger cycloid wavelengths, reaching a maximum at $n$=36 (10$^{\circ}$). In addition, we also obtain that the DMI almost vanishes when the spiral period is $n$ = 3 (120$^{\circ} $).\par

\begin{figure}[htbp]
    \centering
    \includegraphics[scale=0.3]{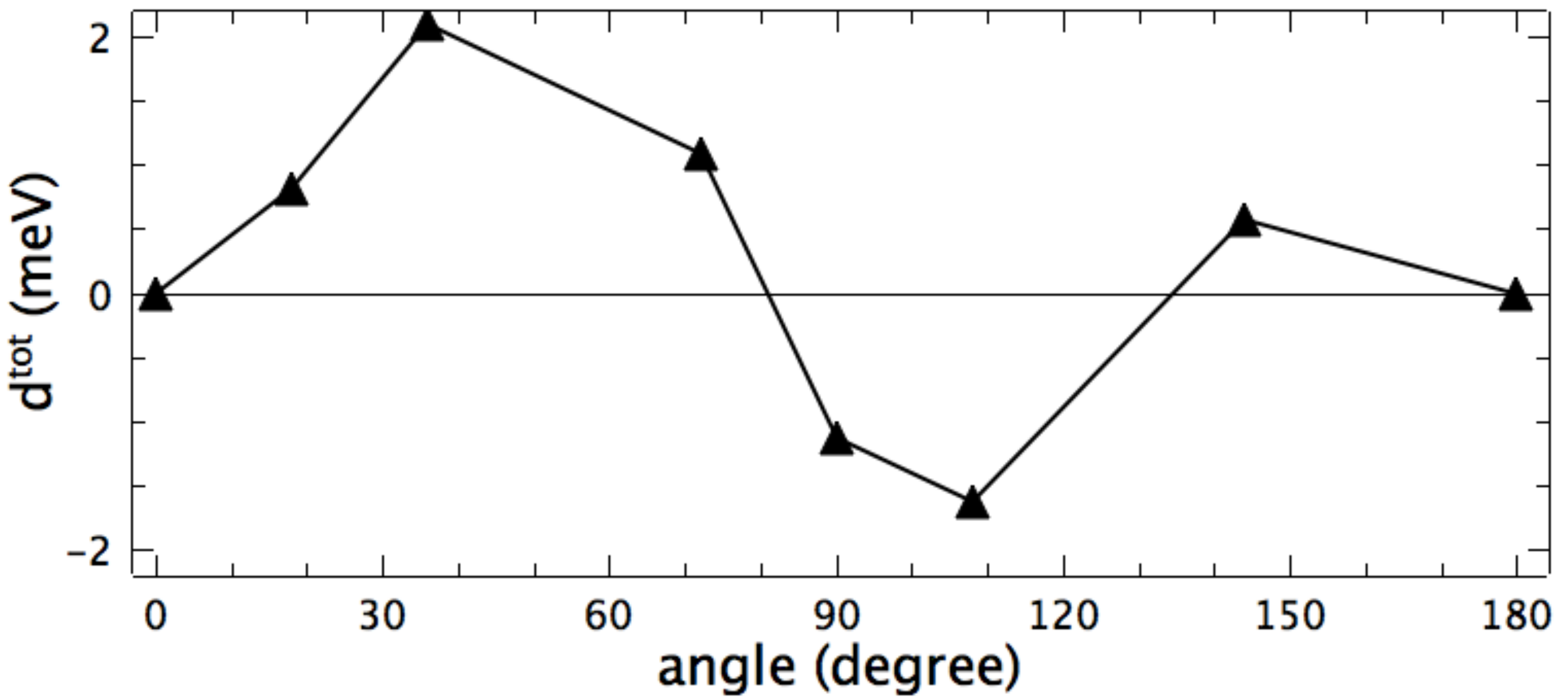}
    \caption{\label{fig6} (Color online) Total DMI energy as a function of the nearest neighbor angle computed using the generalized Bloch theorem method. Although the overall magnitude of $d^{tot}$ is comparable to that obtained using the constrained moment method, the oscillatory behavior indicates that interactions beyond the nearest neighbor approximation are important.}
\end{figure}

For the sake of comparison, we computed DMI as a function of the spin spiral angle using the generalized Bloch theorem, as described in Section \ref{s:method}. For this calculation, we made sure that the structure of the supercell is the same as the one used in the constrained spin spiral calculations. Figure \ref{fig6} displays the angular dependence of the total DMI energy. The DMI now displays an oscillating behavior that is absent in the constrained moment calculations. This oscillatory behavior is not unusual and has been obtained in other transition metal systems such as Fe/Ir(100) surfaces\cite{Menzel2012} or Fe/Pt, Co/Pt zigzag chains \cite{Kashid2014}. It reveals the importance of magnetic interactions {\em beyond} the nearest neighbor approximation, as discussed below. The maximum DMI energy, obtained for an angle of $36^\circ$, is about 2 meV, comparable to the value obtained from the constrained moment calculations at this angle. However, we do not obtain the spike at small angles, nor the extinction at $120^\circ$. 

\subsection{Discussion\label{s:discu}}

In summary, we have shown that the two {\em ab initio} methods (benchmarked against results obtained by other groups) provide very different results, depending on the wavelength considered. Our thickness-dependence study showed that the total DMI interaction receives contributions from Pt monolayers far away from the interface, up to the 4th neighbor (see Fig. \ref{fig3}), emphasizing the importance of intermediate-range interactions beyond the nearest neighbor. In addition, the oscillatory behavior obtained in Fig. \ref{fig6} confirms that DMI energy is not simply proportional to $\sim{\bf S}_i\times{\bf S}_{i+1}$. This deviation from the standard sinusoidal dependence is consistent with Kashid et al.'s finding\cite{Kashid2014}. In their work, the authors stated that the oscillation of the DM energy as a function of the wavelength stems from magnetic interactions beyond the nearest neighbor. This long-range character of the magnetic interactions is specific to metallic transition metal multilayers where the magnetic moment occupies itinerant 3$d$ orbitals. In other words, spin-polarized electrons can convey the magnetic information over several unit cells, and consequently the DMI energy depends on the wavelength of the magnetic structure. As a result, the generalized Bloch theorem is likely to be the most appropriate method to compute DMI in transition metal systems.\par

In contrast, insulating magnets such as MgCr$_2$O$_4$\cite{Xiang2011} or Cu$_2$OSeO$_3$\cite{Yang2012b} involve localized moments interacting via short range interactions (nearest neighbor and next-nearest neighbor hopping). In this case, the DMI energy is defined locally, i.e. the DM vector ${\bf D}_{ij}$ does not depend on the spin spiral wavevector. Then, Eq. (\ref{eq1}) is well justified and the DM energy is expected to display a sinusoidal dependence as a function of the angle between neighboring moments, $\sim\sin\frac{2\pi}{n}$. We point out that Moriya's and Fert's theories of DMI\cite{Moriya1960,Fert1980} both assume localized magnetic moments, and obtain a DM energy that only depends on the angle between neighboring moments. As a result, the constrained moment method is justified to model weak and insulating magnets.

\section{Conclusion\label{s:concl}}
The objective of the present work was to evaluate the applicability and robustness of the constrained moment method, compared to the generalized Bloch theorem approach. We showed that long range interactions, beyond the nearest neighbor, are required to properly account for DMI in transition metals. In particular, we found major discrepancies between the two methods when varying the spin spiral wavelength, which we attribute to the neglect of these long range interactions in the constrained moment method. As a result, the generalized Bloch theorem approach is adapted to itinerant magnets, such as transition metals, while the constrained moment approach is mostly applicable to weak or insulating magnets with localized moments.

\acknowledgments
The research reported in this publication was supported by funding from King Abdullah University of Science and Technology (KAUST). A. M. and A. B. warmly thank M. Chshiev, S. Bl\"ugel and G. Bihlmayer for inspiring discussions.
\bibliographystyle{apsrev4-1}
\bibliography{biblio-full}

\end{document}